\documentclass{elsart}
\usepackage{natbib}
\usepackage{graphicx}
\begin{document}

\begin{frontmatter}
\title{Refracting profiles and generalized holodiagrams}
\author[Criado]{C. Criado},
\author[Alamo]{N. Alamo},
\author[Rabal]{H. Rabal}

\address[Criado]{Departamento de Fisica Aplicada I, Universidad de Malaga, 29071 Malaga, Spain}
\address[Alamo]{Departamento de Algebra, Geometria y Topologia, Universidad de Malaga, 29071 Malaga, Spain}
\address[Rabal]{Centro de Investigaciones Opticas (CONICET-CIC)
P.O.Box 124, 1900, La Plata, Argentina; also with OPTIMO,  Depto.
de Fisicomatem\'aticas, Facultad de Ingenier\'{i}a, Universidad
Nacional de La Plata, Argentina}

\begin{abstract}

The recently developed concept of refracting profiles and that of
refraction holodiagrams are combined so that the classical
Abramson holodiagrams can be generalized taking into account a
wider class of wave fronts and refraction at an interface,
whenever regions of caustics are avoided. These holodiagrams are
obtained as envelopes of specific families of Cartesian Ovals with
an appropriate parametrization. Classical and reflecting
holodiagrams are particular cases of this class. Several of the
properties of the classical holodiagrams are shared by their
richer generalized versions. \vskip.5cm
 \noindent {\it PACS:\/ }
42.40.-i, 42.40.Jv, 42.15.Dp.
\end{abstract}

\end{frontmatter}

\section{Introduction}
In 1969 and 1970, Abramson
\cite{abramson1,abramson2,abramson3,abramson4} proposed the use of
a diagram that condensed many useful properties of holographic
registers, initially concerning the optimal use of available
coherence length and the interpretation of the fringes obtained in
holographic interferometry \cite{abramson5,abramson6}. It
consisted in a family of ellipses showing the loci of equal
optical path length between their common foci and with an adequate
parameterization: consecutive ellipses differed in integer steps
of half the wavelength in the optical path length. The foci
represented a light emitting point source and an observation
point. The chosen parameterization made interpretation of its cuts
as Fresnel zone plates and its volumetric regions as Bragg
gratings. Fermat´s stationary phase principle was also built in
and permitted elementary ray tracing. Numerous additional and non
trivial uses have been found afterwards for that diagram, called
Abramson holodiagram (HD). They include interferometry sensitivity
evaluation, light-in flight registers, Doppler velocimetry, Bragg
diffraction, interpretation of relativistic effects and several
others.

The concept of Abramson holodiagram, originally developed to
describe free propagation in an isotropic medium, has been
broadened to include virtual sources \cite{rabal5}, refraction
between two isotropic media \cite{rabal6,rabal7}, free propagation
in birefringent media \cite{rabal8} and some geometrical aspects
of spatial coherence \cite{rabal9}.

On the other hand, the concept of reflecting profiles
\cite{criado10,criado11} has
 been developed to generate surfaces that modify a spherical
wave front, giving place to another previously determined wave
front or, conversely, to focus a predetermined wave front into a
point. In a recent work \cite{criado12} we have proposed that the
Holodiagram concept can be generalized with the aid of the
reflecting profiles concept. A generalized holodiagram can then be
constructed as a family of reflecting profiles with an adequate
parametrization. It inherits several of the properties of other
holodiagrams defined before.

Recently a method has been proposed to design suitable refracting
profiles (interfaces between two media with different refractive
index) for two different problems: to produce a given wave front
from a single point source after refraction on it, and to focus a
given wave front on a chosen point after refraction (see
\cite{criado13}). In the present paper, the refracting profile
concept is included into the holodiagram construction, giving rise
to the concept of generalized refracting holodiagram (GRHD), which
includes all the (isotropic) holodiagrams previously developed.

The paper is organized as follows. In section 2 we use
\cite{criado13} and construct two families of refracting profiles
by using the optical properties of Cartesian ovals
\cite{lockwood14}. Each profile is constructed as the envelope of
a specific familiy of Cartesian ovals, and depends of a parameter
$a$ that corresponds to half the value of the optical path between
the foci of the family of Descartes ovals used in its
construction.

 In section 3 we use the two families of refracting profiles of
section 2 to construct two kinds of generalized refracting
holodiagrams: one when both source and image are real, and the
other when the source is virtual (or the image is virtual). To
this end we discretize the values of the parameter $a$ in
multiples of half the wavelength employed for illumination.

In section 4 we give some illustrative examples. Also in this
section we use the intersection of the surfaces of the GRHR with
any glass plate surface to design diffraction gratings that work
as generalized Fresnel zone plates. Finally we show how to use the
refracting profiles to design lens that maps a predetermined wave
front $W_1$ into another also predetermined wave front $W_2$.

\section{Refracting profiles}

 \begin{figure}
\begin{center}
\includegraphics{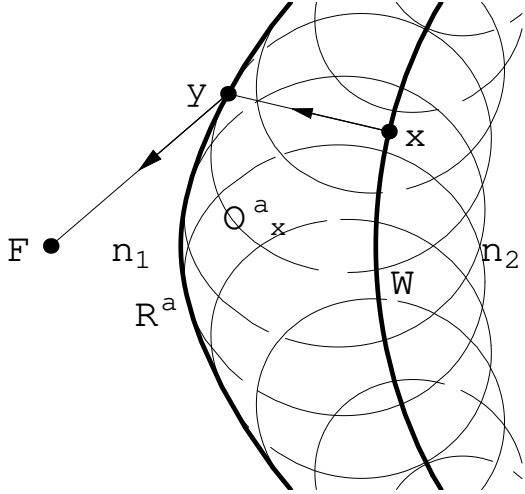}
\caption{The profile $R^a$ is the envelope of a family of interior
Descartes ovals $O_x^a$ with foci $F$ an $x\in W$, where $2a =
n_1\vert y-F\vert + n_2 \vert x-y\vert$. The wave front $W$
focuses in $F$ after refraction at the profile $R^a$.}
\end{center}
\label{Fig1}
\end{figure}

Given a wave front $W$ and a source point $F$, it is possible to
construct a family $\{ R^a\}$ of refracting profiles parametrized
by a non-negative real number $a$, with the property of producing
normal rays to $W$ after refracting the rays that emerge from $F$
at each profile. If light propagation direction is reversed, then
the same family of profiles has the property of focusing the wave
front $W$ in the point $F$. From now on we suppose that the
refracting profile separates two media with refractive indices
$n_1$ and $n_2$, $F$ being in the region of $n_1$, and $W$ in the
region of $n_2$.

The construction of the profile $R^a$ for the wave front $W$ and
the point $F$ proceeds (see \cite{criado13}) by taking the
envelope of a family of Descartes ovals of revolution $\{ O_x^a\}
_{x\in W}$, where $O_x^a$ has foci $F$ and $x$, $x$ varies in $W$
and the parameter $a$ is such that $2a$ is the stationary optical
path length between $F$ and $x$ (see Fig.~1). Let us now recall
some properties of the Cartesian ovals.

The ovals of Descartes or Cartesian ovals were introduced by
Descartes in 1637 in his Dioptrique, dedicated to the study of
light refraction. A Cartesian oval is the locus of the points from
which the distances $r_1$ and $r_2$ to two fixed points $F_1$ and
$F_2$, called foci, verify the bipolar equation
\begin{equation}\label{oval}
a r_1 + b r_2 = k,
\end{equation}
where $a$, $b$ and $k$ are constants. Observe that this equation
includes the bifocal definition of conics for the particular cases
of $k>0$ and $a=b>0$  (ellipse), or  $a= -b>0$ (hyperbola). A
Cartesian oval has a third focus $F_3$, and the oval can be
defined by any two of the foci (see \cite{lockwood14}). In
particular when $\vert a/b\vert$ goes to 1, the third focus goes
to infinity and we get the conics.

The so-called complete Cartesian oval is the set of curves
associated to the bipolar equation
\begin{equation}\label{completeoval}
a r_1 \pm b r_2 = \pm k \quad (k>0)
\end{equation}

 \begin{figure}
\begin{center}
\includegraphics{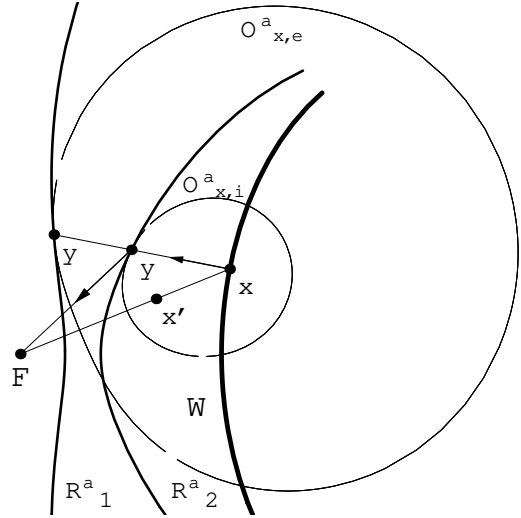}
\caption{The wave front $W$ and the two profiles $R_i^a$,
$i=1,2)$, which are the envelopes of the interior ($O_{x,i}^a$)
and the exterior ($O_{x,e}^a$) Descartes ovals. These ovals have
foci $F$ and $x\in W$; $x'$ is the third focus. The wave front $W$
focuses in $F$ after refraction at the profile $ R_2^a$.}
\end{center}
\label{Fig2}
\end{figure}

Only two of the four equations obtained from these double signs
are not empty. They are closed curves, one interior to the other
(see Fig.~2 with $F_1 = F$ and $F_2 = x$). These two curves
intersect only when two foci coincide. In this case we get the
so-called Lima\c con of Pascal. See \cite{lockwood14} for a
detailed study of these ovals. In the context of the geometrical
optics, the coefficients $a$ and $b$ correspond to the refractive
indices, so that equation \ref{oval} gives the stationary optical
path length between the foci $F_1$ and $F_2$.

In what follows we will use complete Cartesian oval of revolution,
that is the surfaces obtained when the above ovals are rotated
around their focal axes. Let $O_x^a$ denote the complete Cartesian
oval of revolution with foci $F$ and $x\in W$, and let $O_{x,i}^a$
(respectively $O_{x,e}^a$) denote the interior (respectively
exterior) oval. If we assume that $F$ is the origin, then for any
points $x\in W$ and $y\in O_x^a$, the vectors $Fx$ and $Fy$ with
origin $F$ and extreme $x$ and $y$ respectively, will be denoted
again by $x$ and $y$. By the definition of cartesian Ovals these
vectors verify:
\begin{equation}\label{concreoval}
n_1 \vert y\vert \pm n_2\vert y-x\vert = \pm 2a.
\end{equation}

The interior oval $O_{x,i}^a$ verifies $n_1 \vert y\vert +
n_2\vert y-x\vert = 2a$, and the exterior oval $O_{x,e}^a$
verifies $- n_1 \vert y\vert + n_2\vert y-x\vert =  2a$ (see
\cite{lockwood14}).

Let $R^a$ be the envelope of the family $\{ O_a^x\} _{x\in W}$. To
give an explicit parametrization of the profiles $R^a$ let $u(x)$
be an unitary normal vector to $W$ at $x$. A point $y\in R^a$ can
be given by the parametric equation:
\begin{equation}\label{eqnormal}
y= x + \lambda (x) u(x),
\end{equation}
where $\lambda (x)$ has to be determined by the condition that
$y\in O_x^a$, that is, $y$ has to verify equation
\ref{concreoval}.

 A straightforward calculus gives two $\lambda$'s, when the
  profiles and $F$
are at the same side with respect to $W$:
\begin{equation}\label{eqlambdas}
\lambda_{i} = {{2 a n_2 - (-1)^i n_1^2 (x\cdot u) + (-1)^{i}
\sqrt{\Delta_i} }\over {n_2^2 - n_1^2}}, \quad  i= 1,2,
\end{equation}
which are defined only when $\Delta_i = (2 a n_2 - (-1)^i n_1^2
(x\cdot u))^2 - (n_2^2 - n_1^2)(4 a^2 - n_1^2 x^2)$ is positive.
Then we obtain two sheets $R_1^a$ and $R_2^a$  parameterized by
\begin{equation}\label{eqnormali}
y= x + \lambda_{i} (x) u(x), \quad  i= 1,2.
\end{equation}
These two sheets are the envelopes of the exterior and interior
ovals respectively. Figure 2 illustrates these sheets in a
2-dimensional example.

 The profiles $R^a$ are smooth surfaces except
for the points $y\in R^a$  that are centers of curvature of the
wave front $W$, so profiles with singularities can only appear for
concave wave fronts. The geometrical locus of the centers of
curvature of $W$, or equivalently, the envelope of the normal
lines to $W$, is usually called the caustic, $C$, of $W$. The
singularities of the profiles $R^a$ sweep out also the caustic as
the parameter $a$ varies (see \cite{criado13}).

We should observe that if we want that the profiles have physical
sense, we have to exclude the points $y$ of the profiles such that
the corresponding incidence angle $\theta_1$ does not give a real
value for the angle of refraction $\theta_2$. It occurs when light
is propagated from an optically denser medium into one that is
less optically dense, i.e. when $n_2< n_1$, provided that the
incidence angle $\theta_1$ exceeds the critical value $\theta_c$
given by  $\sin \theta_c = n_2/n_1$.

\section{Generalized Refraction Holodiagrams}
Using the profiles $R^a$ we can construct generalizations of both
kinds of refraction holodiagrams: one when both source and image
are real \cite{rabal6}, and the other when the source is virtual
(or the image is virtual) \cite{rabal7}. The sheets $R_2^a$
correspond to generalizations of the first case and the sheets
$R_1^a$ to the second. This is because for the particular case of
$W$ being a spherical wave front, $R_2^a$ (respectively $R_1^a$)
is an interior (respectively exterior) Cartesian oval as is
required in the first and second cases respectively.

a) First case.

 To construct the holodiagrams we take the
family of profiles $R_2^a$ given by equation \ref{eqnormali} and
$i =2$, with the parameter $a$ varying in halves of the
wavelength, so that $R_2^a$ is the envelope of a family of
interior Cartesian ovals. From the construction of $R_2^a$, any
$y\in R_2^a$ verifies $n_1 \vert y\vert + n_2\vert y - x\vert =
2a$, where $x\in W$ is in the normal line from $y$ to $W$. In this
case, (see Fig.~2), if $a$ is such that $F$ and $W$ are at
different sides of the tangent plane to the profile at any point
of it, then $R_2^a$ focuses $W$ in $F$ ($R_2^a$ is a refracting
focusing profile); or reversing the ray direction, $W$ is
reconstructed from light coming from the source point $F$ ($R_2^a$
is a refracting profile).

We have called this holodiagrams Generalized Abramson Holodiagrams
(GAHD) because for $n_2 = n_1$ and $W$ a spherical wave front we
obtain the classical Abramson Holodiagrams. Note also that if $n_2
= n_1$ and $W$ is any arbitrary wave front we obtain the
Reflective Holodiagrams of \cite{criado12}.

b)  Second case.
 \begin{figure}
\begin{center}
\includegraphics{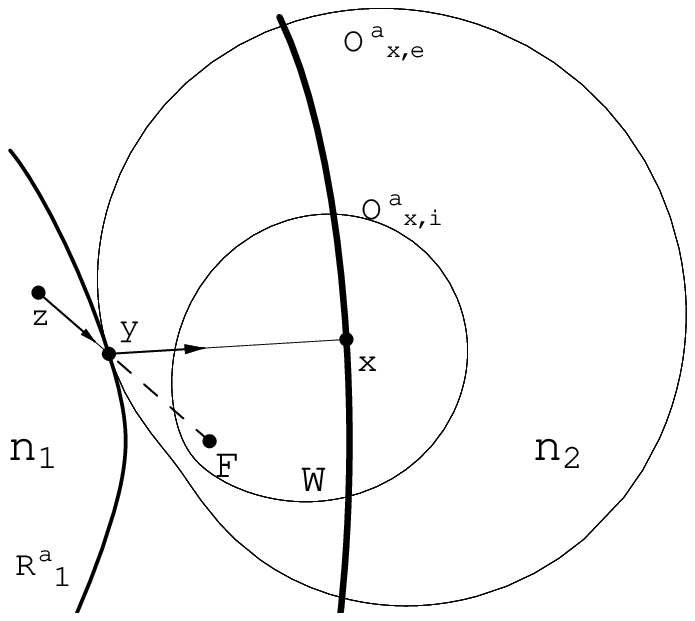}
\caption{For $F$ in the interior of the ovals, a spherical wave
front converging to $F$ produces, after refraction at$ R_1^a$, the
wave front $W$.}
\end{center}
\label{Fig3}
\end{figure}

In this case we take the family of profiles $R_1^a$ given by
equation \ref{eqnormali} and $i = 1$, with the parameter $a$
varying in halves of the wavelength, so that $R_1^a$ is the
envelope of a family of exterior Cartesian ovals. As before, from
the construction of $R_1^a$, any $y\in R_1^a$  verifies $- n_1
\vert y\vert + n_2\vert y - x\vert = 2a$, with $x\in W$ in the
normal line from $y$ to $W$. In this case, (see Fig.~3), if $a$ is
such that $F$ and $W$ are at same side of the tangent plane to the
profile at any point of it, then we have that any spherical wave
front $S$ converging to $F$ from the medium of $n_1$, after
refracting at $R_1^a$ produces the wave front $W$. Equivalently,
we have that the wave front $W$ after refracting at $R_1^a$
produces a spherical divergent wave front $S$ with center $F$. To
verify this consider a point $z\in S$, so it satisfies $\vert
z\vert = R$(constant); then the optical path length $n_1 \vert z -
y\vert + n_2\vert y - x\vert = 2a$ from $y\in W$ to $z\in S$ must
be constant. Since $\vert z - y\vert = R -\vert y\vert$, then $-
n_1 \vert y\vert + n_2\vert y - x\vert$ must be constant, and this
is just the equation that the points $y$ in the exterior ovals
satisfy, whose envelope is $R_1^a$. We have called these
holodiagrams Generalized Young Holodiagrams (GYHD) because for
$n_2 = n_1$ and $W$ a spherical wave front we obtain the Young
Holodiagrams.

\section{Results and examples}
We show now some illustrative examples of both kinds of
Generalized Refracting Holodiagrams obtained with the described
procedure.

The images are shown as gray levels proportional to $1 + \cos (r
a)$, with $r$ a scale constant, so that they depict as fringes the
families obtained when the parameter $a$ is increased in a
continuous way.

The shape of the fringes indicates the loci of stationary optical
path sum or difference and the geometry of a diopter such that
when illuminated by a point source at $F$ generates a wave front
in the shape of $W$. The orientation and spacing of these fringes
indicates the sensitivity to phase variations. Places where
fringes are dense indicate that phase changes fast there with
position, while the direction of the fringes indicates the zero
sensitivity direction. Loci of equal sensitivity, that is the $K$
curves defined by Abramson, can be numerically determined from
these figures. Phase changes or departure of the calculated values
could be due, for example, to construction errors in the
refracting surfaces. Then, the spacing between the fringes is an
estimation of its local construction tolerance.

 \begin{figure}
\begin{center}
\includegraphics{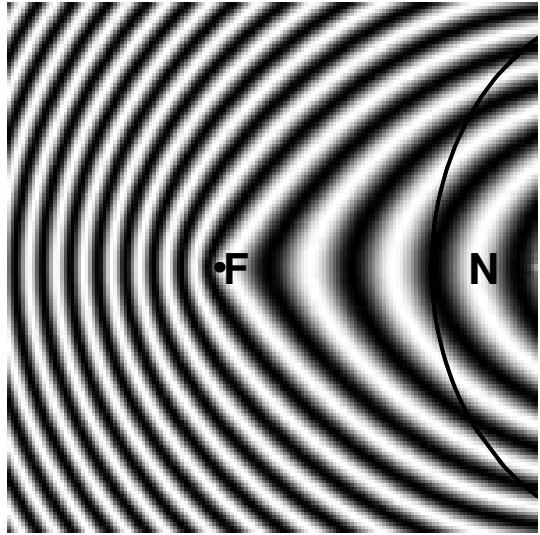}
\caption{Generalized Abramson holodiagram GAHD for a convex wave
front $W$ and focus $F$.}
\end{center}
\label{Fig4}
\end{figure}

 \begin{figure}
\begin{center}
\includegraphics{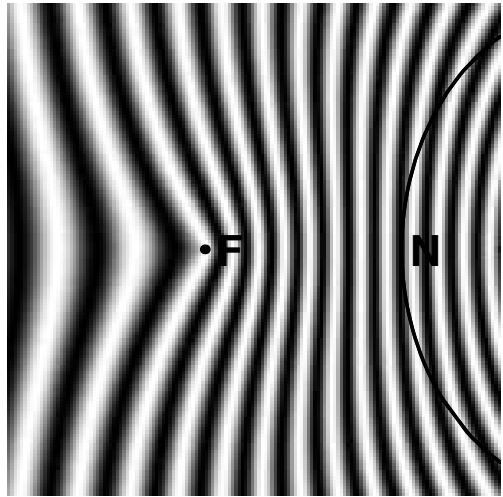}
\caption{Generalized Young holodiagram GYHD for a convex wave
front $W$ and focus $F$.}
\end{center}
\label{Fig5}
\end{figure}

Figures 4 and 5 show the generalized Abramson (GAHD) and Young
(GYHD) holodiagrams for a convex elliptical wave front.  Figures 6
and 7 show the generalized Abramson (GAHD) and Young (GYHD)
holodiagrams for a wave front with sinusoidal form.

\begin{figure}
\begin{center}
\includegraphics{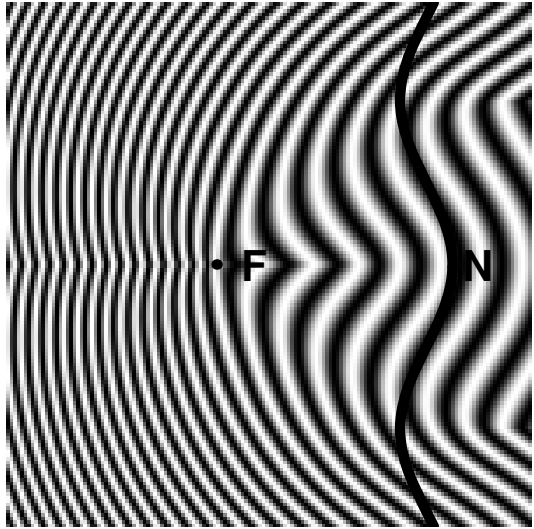}
\caption{Generalized Abramson holodiagram GAHD for a wave front
$W$ with sinusoidal shape and focus $F$.}
\end{center}
\label{Fig6}
\end{figure}
\begin{figure}
\begin{center}
\includegraphics{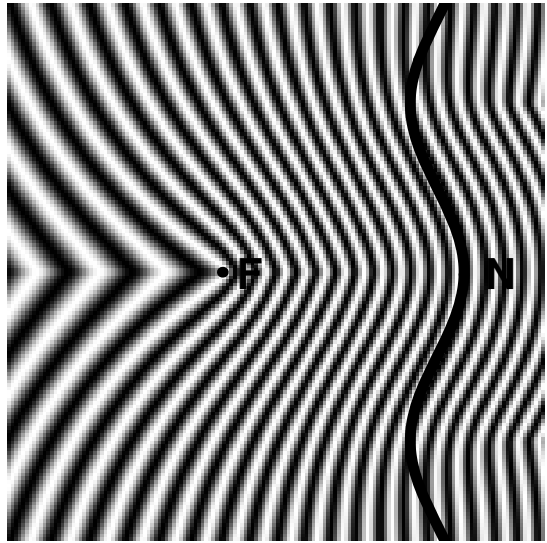}
\caption{Generalized Young holodiagram GYHD for a wave front $W$
with sinusoidal shape and focus $F$.}
\end{center}
\label{Fig7}
\end{figure}

In a similar way as in other holodiagrams, a cut of these diagrams
by an arbitrary surface defines a generalized Fresnel Zone Plate.
The now known as Fresnel Zone Plate was invented by Lord Rayleigh;
he wrote in 1871: "The experiment of blocking out the odd Huygens
Zones  so as to increase the light of the center succeeded very
well..." \cite{jenkins}. A classical Fresnel Zone Plate is a
diffraction gratings device that consists in a binary pupil that
is transparent in alternate Fresnel zones and absorbing in the
rest. When it is illuminated by a spherical wave front,
constructive interference occurs in certain points of its axis and
high intensity is observed there, so that these points can be
considered as images of the source. For each image point it can be
considered as a device that maps a spherical diverging wave front
from the source into a spherical converging wave front to the
image or as the hologram of a point, being all these descriptions
equivalent. We  are going to call a Generalized Fresnel Zone Plate
(GFZP) to a pupil that diffracts a predetermined wave front when
illuminated by a given one, in our case a spherical diverging wave
or, conversely it maps a given wave front into a spherical
converging wave. The GFZP is indeed a hologram and the new name is
assigned to it only to help in its interpretation.

Now, let us assume, as Abramson does, that the GRHD has the
imaginary property to paint the surfaces it intersects with its
local gray level. For easy description let us also assume that we
use a binary version of the GRHD, that is, we use the
approximation that the sine wave of light is just a square wave
that has one half in antiphase with the other. Then, when the GRHD
cuts any arbitrary shaped surface $\Sigma$, limiting the media
with the refractive indices used in its calculation, a series of
painted and transparent regions are defined on the surface. If the
so painted $\Sigma$ surface is now illuminated by a point source
at F it will be transmitted only in the transparent regions. That
is to say that only points fulfilling the condition that the sum
of the optical paths between $F$ and tha wave front $W$ differ in
integer multiples of $\lambda$, will contribute to the transmitted
wave front. The painted surface behaves as a hologram that when
illuminated by a wave front coming from $F$ diffracts a wave front
congruent with $W$ and conversely. We can say then that such a cut
is a Generalized Fresnel Zone Plate.

It can be easily seen  that, if the refracting surface is itself a
refracting profile, then the Fresnel Zone plate is not needed: the
surface is uniformly transparent. Then, if a certain refracting
surface shape is close to that of a refracting profile, this
Fresnel Zone plate or hologram will show extensive uniform
regions. This means that the Fresnel Zone plate so generated can
be used to correct construction phase errors in a refracting
profile and it will consist in a low spatial frequency register.

Finally we will give another optical application of the refracting
profiles construction: As, when the incident wave front is $W$,
after refraction in the profile the wave front is spherical
converging to or diverging from $F$, the second medium can be
finished as a spherical surface centered in $F$ \cite{rabal6}
followed by the same refracting medium with index $n_1$ or any
other. Then, the focusing profile becomes a lens that either
selectively focuses a given wave front $W$ or synthesizes from $F$
a predetermined wave front $W$ in the same medium. As shown in
reference \cite{rabal7}, the second surface finishing needs not to
be exclusively spherical but can also be in the shape of a
Cartesian oval. Moreover, as the refracted wave front is spherical
converging to $F$, the same concept described here permits to
design the second  surface as a second refracting profile. The
composite effect of the two cascaded refracting profiles acts then
as a particular lens that maps a predetermined wave front $W_1$
into another also predetermined wave front $W_2$ after two
refractions.

Conversely, a spherical wave front, either diverging from or
converging to a point $F$, can be mapped with a refracting profile
into another wave front and the latter again mapped after a second
refraction to a diverging or converging wave front  with the same
or other focus $F$. The composite effect is that of a sort of lens
that conjugates a point source to the final focus with an
encryption step between. The generalization of the HD to two more
arbitrary wave fronts will be treated in a forthcoming work.

\section{Conclusions}
We have proposed a generalization of the classical concept of
Holodiagram with respect to one wave front. It is based on the
concepts of refracting and focusing profiles recently developed.
Regions corresponding to caustics and auto intersections need to
be excluded. We have shown some characteristic examples.

The use of the new compound concept can be found not only in
getting insight about wave front propagation but also in the
design of computer generated holograms to synthesize predetermined
wave fronts, in pattern recognition of specified wave fronts, in
the mapping of a wave front into another, in the calculation of
tolerances in special lenses construction, in elementary ray
tracing, encryption, etc.

\vskip1cm \noindent
 {\bf Acknowledgements}

This work was partially supported by Spanish grants FQM-192 (C.
Criado) and BFM2001-1825 (N. Alamo), and CONICET and Faculty of
Engineering, Universidad Nacional de La Plata, Argentina (H.
Rabal).

\end{document}